\documentclass[oribibl]{llncs}

\usepackage[english]{babel}
\usepackage[latin1]{inputenc}
\usepackage{amsfonts}
\usepackage[cmex10]{amsmath}
\usepackage{url}
\usepackage{graphicx}
\usepackage{clrscode}
\usepackage{subfig}

\usepackage{makeidx}  

\newenvironment{keywords}{
       \list{}{\advance\topsep by0.35cm\relax\small
       \leftmargin=1cm
       \labelwidth=0.35cm
       \listparindent=0.35cm
       \itemindent\listparindent
       \rightmargin\leftmargin}\item[\hskip\labelsep
                                     \bfseries Keywords:]}
     {\endlist}

\begin{document}
\frontmatter          
\pagestyle{headings}  
%
%
%
\mainmatter              
\title{Recommender Systems by means of Information Retrieval}
%
%
\author{Alberto Costa\inst{1} \and {Fabio Roda}\inst{2}}
%
%
%
\institute{LIX, \'Ecole Polytechnique, F-91128 Palaiseau, France\\
\email{costa@lix.polytechnique.fr}
\and
LIX, \'Ecole Polytechnique, F-91128 Palaiseau, France\\
\email{roda@lix.polytechnique.fr}
}

\maketitle              

\begin{abstract}
  In this paper we present a method for reformulating the \mbox{Recommender} Systems problem in an Information Retrieval one. 
  In our tests we have a dataset of users who give ratings for some movies; we
hide some values from the dataset, and we try to predict them again using its remaining portion (the so-called ``leave-n-out approach'').

In order to use an Information Retrieval algorithm, we reformulate this Recommender Systems problem in this way: a
user corresponds to a document, a
  movie corresponds to a term, the active user (whose rating we want to predict)
plays the role of the query, and the
  ratings are used as weigths, in place of the weighting schema of the original
IR algorithm.

The output is the ranking list of the documents (``	users'') relevant for the
query (``active user''). We use
  the ratings of these users, weighted according to the rank, to predict the rating of
the active user.  
  We carry out the comparison by means of a typical metric, namely the accuracy
  of the predictions returned by the algorithm, and we compare this to
  the real ratings from users.  
  In our first tests, we use two different Information Retrieval algorithms: LSPR, a recently proposed model based on Discrete Fourier Transform, and a simple vector space model.
\end{abstract}


\section{Introduction}
Recommender Systems (RS) have become an important research area and,
as a consequence, many empirical studies appeared in recent years.
Many methods have been proposed by different, important research groups and,
in particular, Collaborative
Filtering~\cite{SchaferFrankowskiEtAl07p291,Sarwar-2001, VozalisMargaritis03,
Linden-et-al-2003} gained
a great popularity and it is nowadays a well known approach.  
It was remarked \cite{Belkin92informationfiltering} that collaborative filtering
shares fundamental aspects with
Information Retrieval (IR) and there is somehow a continuity between these two fields of
research. 
This work belongs to this stream. In fact we are developing a RS
which makes 
use of concepts and tools used elsewhere in an IR
context and, believing
that 
the underlying structure could also provide
an interesting framework for RS algorithms, we looked
for experimental evidence of this intuition. 

In our approach first we move from the Recommender Systems domain to the Information Retrieval one. To do this, we consider each user as a document and each movie as a term (even if we can use this approach not only for the movies): in this way, as in IR a document is a set of terms, in the RS field a user becomes as a set of movies (for which the user has given a rating). Using this representation, the ratings of the users are the baseline for computing the weights of the terms, as explained in section \ref{sec:background}.

Moreover, the active user becomes the query in this phase; the meaning of this is that in Information Retrieval we want the documents more similar to the query, and for the RS problem we want the users more similar to the active user.

At this point we can use one of the several existing IR algorithms to obtain the ranking list, that represents the set of users more similar to the active user, ordered by decreasing similarity.

Finally, as explained at the end of section \ref{sec:background}, we use the ranking list to get the predicition for the active user; this last step brings us again in the Recommender Systems domain.

The evaluation of most works in this field is carried out using
``artificial" datasets   
provided by well know research groups, such as GroupLens \cite{229,227}, or by
Netflix (\url{http://www.netflixprize.com}) \cite{1669716, b257}. This approach
ensures somehow a standard method to evaluate results.
Hence, we have implemented our in-house algorithm using both Least Spectral Power
Ranking (LSPR) model, presented in~\cite{Lspr}, and an algorithm based on vector space model, as conceived in the 1960s by Salton ~\cite{Salton68,Salton79}; we tested it by means of a
standard dataset provided by GroupLens. Basically, we have compared it 
with the ``community'' which constitutes the benchmark to overcome, in order to show the feasibility of the approach.

This paper is organised as follows. In section~\ref{sec:problem} we introduce the
problem in a more
formal way, and in section~\ref{sec:background} we describe the 
algorithm itself. After that, in
section~\ref{sec:testing} we report the experimental results obtained
when running this algorithm. Finally, in
section~\ref{sec:conclusion} we discuss the results.

\section{Description of the problem}\label{sec:problem}

We can formulate our problem as follows. 
We have: 
\begin{itemize}
        \item a set $U$ of users, $|U| = n$;
        \item a set $I$ of items (movies, songs, restaurants...), $|I| = m$;         
        \item a gain function G which expresses the utility of an item for a user
\end{itemize}
Utility is expressed by a numeric value representing a rating (the higher the
better) varying on a chosen
interval V, more formally the function G is defined as:
\begin{center}
G: $U$x$I$ $\rightarrow$ V
\end{center}
We want to maximize the users' utility by recommending good items and
advising against bad ones.
The problem is that we do not know ``a priori'' all the values of G, hence we have to
predict users' ratings.
This ability is normally tested in an almost empirical way showing that the system is
able to predict a set of known ratings.

In this paper we use a dataset provided by
GroupLens. Basically
there are 100 000 ratings (from 1 to 5) given by $n=943$ users on $m=1682$ movies and 
 each user rated at least 20 movies.
We represent this dataset with a matrix $D^{m\text{x}n}$, where $D_{ij}$ is the
rating given by the user $j$ for the movie $i$ (0 value is used if no rating is
available).
Our aim is to predict the rating of a user (called ``active user'') for each movie
(using the informations of the matrix $D$), minimizing the differences between the
predicted ratings and the real ones.

\section{Model}\label{sec:background}
This section describes the core concepts of our framework, where we use both the LSPR model and the vector space model as IR algorithms. 

Basically in the LSPR model the query is viewed as a spectrum and each document as a
set of filters, with one filter for each document term, whereas the vector space model views terms as basis vectors, documents and queries as vectors of
the same space.

Usually in Information Retrieval some weighting schemes are used for the terms of the documents and for the terms in the query; the basic choice is to use the TF-IDF weighting schema for the former, and the IDF weighting schema for the latter. In
order to use the IR algorithms for the Recommender Systems, it is necessary to modify these
weighting functions.
Since each user becomes a document, and each movie becomes a term, there is a
similarity between the matrix $D$ and the well-known term-document matrix.
At this point, consider an active user $k\in U$, for which we want to predict the
rating for the movie $h\in I$. Starting from $D$, we compute a new matrix $WU$ (that
plays the role of the normalized TF-IDF weights matrix in Information Retrieval) as
follows:
\begin{equation}
\label{eq:DW}
WU_{ij}=
\begin{cases}
0 & \text{if } D_{ij}\cdot D_{ik}=0\\
1-\frac{|D_{ij}-D_{ik|}}{4} & \text{otherwise}
\end{cases}
\end{equation}
This means that the more the rating of a user for a movie is similar to the rating
of the active user, the more its weight (from 0 to 1).

The column $k$ in this matrix is not considered, because it is 0 for the movies not
rated by the active user, 1 otherwise. 

After that, we compute the weights for the active user (i.e. the IDF weights for the
query); we save these informations in the column $k$ of the matrix $WU$, using the
following formula:
\begin{equation}
\label{eq:QW}
WU_{ik}=
\begin{cases}
0 & \text{if } n_i=0 \text{ OR } D_{ik}=0\\
\log_2\left(\frac{n}{n_i}\right) & \text{otherwise,}
\end{cases}
\end{equation}
where $n_i$ is the number of users that have rated the movie $i\in I$.

Now we are ready to use the Information Retrieval algorithm: the query is represented by the column $k$
of $WU$, while the documents of the collection are the columns $j\neq k$ of the same
matrix with $WU_{hj}\neq 0$. The output of the model is the ranking list of the
documents, ordered by increasing relevance. This means that the collection is the
set of users that have rated the movie $h$, and the output is the same set of users
ordered from the more to the less ``similar'' to the active user.

The last operation is to predict the rating. To do this, we use the ratings of the
users in the ranking list, weighted by their rank, so that the smaller the rank of
the user is, the more his rating is considered.
Suppose the ranking is given by the list of users $R$, where $|R|$ is the number of
retrieved users, the
rank of each user is from 0 (first) to $|R|-1$ (last), and
$D_{h,j(r)}$ is the rating for the movie $h$ of the user with rank $r$. The
predicted rating is computed as:

\begin{equation}
\label{eq:p_hk}
  p_{hk}=\frac{\displaystyle\sum_{r=0}^{|R|-1}{\left(1-\frac{r}{|R|}\right)\cdot
D_{h,j(r)}}}{\lambda},
\end{equation}
where $\lambda$ is the normalization term, computed as
\begin{equation}
\label{eq:lambda}
\lambda=\sum_{r=0}^{|R|-1}{\left(1-\frac{r}{|R|}\right)}=\frac{|R|+1}{2}.
\end{equation}

Figure \ref{fig:palgo} summarizes the algorithm.
Basically, the rows from 1 to 14 represent the operation described by equation
\eqref{eq:DW}, while the rows from 15 to 23 implement the equation \eqref{eq:QW}.
Finally there is the call to the Information Retrieval algorithm, and the prediction of the rating,
according to equations \eqref{eq:p_hk} and \eqref{eq:lambda}.

\begin{figure}[!htb]
  \begin{codebox}
    \Procname{\textbf{Algoritm:} RecSys-to-IR\\
      \textbf{Input:} data set D, active user $k$, movie $h$, IR algorithm \\
      \textbf{Output:} prediction $p_{hk}$}
    \li \For \textbf{each} $i\in I $
    \li \Do
    \li $n_i\leftarrow 0$
    \li \For \textbf{each} $j\in U | j\neq k$ 
    \li \Do
    \li \If $(D_{ij}\cdot D_{ik}=0)$
    \li \Then
    \li $WU_{ij}\leftarrow 0$
    \li \Else
    \li $WU_{ij}\leftarrow 1-\frac{|D_{ij}-D_{ik|}}{4}$
    \li $n_i\leftarrow n_i+1$
    \End
    \li \textbf{end if}
    \End 
    \li \textbf{end for}
    \End
    \li \textbf{end for}
    \li \For \textbf{each} $i\in I $
    \li \Do
    \li \If $(n_i=0)$
    \li \Then
    \li $WU_{ik}\leftarrow 0$
    \li \Else
    \li $WU_{ik}\leftarrow \log(\frac{n}{n_i})$
    \End
    \li \textbf{end if}
    \End
    \li \textbf{end for}
    \li Call the IR algorithm, and get the ranking list $R$
    \li $p_{hk}\leftarrow\text{Round}\left(2\cdot
\frac{\sum_{r=0}^{|R|-1}{\left(1-\frac{r}{|R|}\right)\cdot
D_{h,j(r)}}}{|R|+1}\right)$
    \li \textbf{return} $p_{hk}$
  \end{codebox}
  \caption{The prediction algorithm.}
  \label{fig:palgo}
\end{figure}

\section{Evaluation of the algorithm}\label{sec:testing}

 Many different measures are used in order to evaluate the performance
 of filtering algorithms employed by Recommender Systems and some
 metrics fit better for top-N recommendation, and others for
 prediction  \cite{963772}. We decided to use a simple metric to evaluate our
 system, in order to have clear preliminary results easy to understand.
 We basically used the accuracy
 computed as the square root of the averaged squared difference
 between each prediction and the actual rating (the root mean squared
 error or ``RMSE"). 
 
 Let $p_{ik}$ denote the predictions generated by
 a certain algorithm for a set with $k = 1, 2, ..., |k|$, and let the
 actual ratings provided by a certain user be denoted by $ra_{ik}$  $(k
 = 1, 2, ..., |k|)$.  RMSE is defined by:
\[
\text{RMSE } = \sqrt{\frac{\sum_{k=1}^{|k|} {(ra_{ik} - p_{ik})}^2}{|k|}}
\]

In our tests, we round the $p_{ik}$ values to the closest integer number, because
the real ratings are integers.
As mentioned above, the evaluation of RMSE is typically performed
using the ``leave-n-out'' approach \cite{breesetechrep}, where a part
of the dataset is hidden and the rest is used as a training set
for the Recommender Systems, which tries to predict properly the withheld
ratings. 

We calculate the predictions with 
LSPR and the basic vector space algorithms using data from the training set and we compare
the prediction against the real rating in the test set. 
As a benchmark to evaluate the algorithm we employ the
\emph{community average} for a certain item, with the aim of measuring
how much our algorithm can improve the simple \emph{community
recommendation}. Thus, we also compute the RMSE of the community
recommendation with respect to the actual ratings provided by the users.

The results reported in table \ref{table:all} refer to the analysis we
performed using the dataset from GroupLens\footnote{The dataset can be found on \url{http://www.grouplens.org}.} described above.
We used five couples (training set, test set) which share the same composition
(80\%/20\% spilts of the orginal data into training and test data) as suggested by 
the guidelines of GroupLens itself.

\begin{table}[ht]
\centering
\begin{tabular}{|c|c|c|c|} \hline
set & LSPR & vector space& community \\ \hline
1 & 0.985 & 0.989 & 1.073 \\ \hline
2 & 0.974 & 0.984 & 1.067 \\ \hline
3 & 0.971 & 0.980& 1.060 \\ \hline
4 & 0.967 & 0.979 & 1.056 \\ \hline
5 & 0.975 & 0.984 & 1.065 \\ \hline

\hline\end{tabular}
\caption{RMSE}\label{table:all}
\end{table}

In table~\ref{table:percall} we express
this result in relative terms by providing the \emph{rate of
  improvement} with respect to the average of the ratings by the
community: LSPR overcomes the community by 8.4\% on
average, while the vector space model decreases the RMSE by 7.6\%. 

\begin{table}[ht]
\centering
\begin{tabular}{|c|c|c|} \hline
set& Improvement LSPR & Inprovement vector space\\ \hline
1 & 8.2 \% & 7.8 \% \\ \hline
2 & 8.7 \% & 7.8 \%\\ \hline
3 & 8.4 \% & 7.5 \%\\ \hline
4 & 8.4 \% & 7.3 \%\\ \hline
5 & 8.5 \% & 7.6 \%\\ \hline
\textbf{mean} & \textbf{8.4 \%} & \textbf{7.6 \%} \\ \hline
\end{tabular}\caption{Improvement over community
average.}\label{table:percall} 
\end{table}

\section{Discussion and conclusions}\label{sec:conclusion}

A first consideration is that the algorithm already outperforms the community 
even if the gap is not prodigious. As a matter of fact, other algorithms recently proposed by different authors, like \cite{Kozma_1,lgdea}, show RMSE values in the range 0.88 - 0.95 for the same dataset. However, the aim of this paper is to show that our new approach could be a basis for a more sophisticated algorithm, rather than presenting an algorithm already comparable with the state-of-the-art.

So, first of all, we plan to work on fine-tuning our algorithm, to extend this 
empirical evaluation and to compare 
it with some well known algorithms such as KNN or Slope-one
\cite{LemireMaclachlan2005}.

We also plan to perform 
further experiments 
with a larger number of datasets and with a finer grain analysis of sensitivity
of the algorithm with respect to the size of the data set. In parallel, we
are also working on LSPR to improve its scalability and to include a
number of optimization techniques.

Moreover, we plan to use other weighting schemes, as the well-known Okapi BM25, and other IR algorithms, for example probabilistic models like Terrier \cite{ter1, terrier2}. The final aim of this work is to merge the results to obtain better performances, as already done in Information Retrieval \cite{ChenK06}.

Basically, we are still trying to better understand if our approach can provide a nice outcome
in the Recommender Systems field and
we consider the present work as a first answer, so our contribution is an experimental investigation of some possible relations between
Information Retrieval and Recommender Systems. 
In this sense it seems very interesting that the results obtained with LSPR are better than the ones obtained with the vector space model, reflecting the behaviour of these models in the IR field, as reported in \cite{Lspr}.
\bibliography{recsy}

\bibliographystyle{splncs}

\end{document}